\newif\ifDEBUG

\newif\ifANONYMOUS
\ANONYMOUSfalse

\newif\ifARXIV
\ARXIVtrue

\ifANONYMOUS
  \documentclass[acmsmall,screen,review,anonymous=true,natbib=false]{acmart}
\else
  \ifARXIV
    \documentclass[acmsmall,screen,anonymous=false,natbib=false,nonacm]{acmart}
  \else
    \documentclass[acmsmall,screen,anonymous=false,natbib=false]{acmart}
  \fi
\fi

\AtBeginDocument{%
  }

\ifARXIV
  \setcopyright{cc}
  \copyrightyear{2025}
  \acmYear{2025}
\else
    \setcopyright{cc}
    \copyrightyear{2026}
    \acmVolume{1}
    \acmNumber{FSE}
    \acmArticle{FSEXXX}
    \acmYear{2026}
    \acmMonth{7}
    \acmDOI{10.1145/XXXXXXX}
    \acmJournal{PACMSE}

    \acmISBN{2994-970X/2026/7-ARTFSEXXX}

\fi

\RequirePackage[
  datamodel=acmdatamodel,
  style=acmnumeric,
  ]{biblatex}

\usepackage{misc/acmart-taps}
\usepackage[export]{adjustbox} %
\usepackage{amsmath}
\usepackage{amsfonts}
\usepackage{array}
\usepackage[USenglish]{babel}
\usepackage{misc/bootstrapicons}
\usepackage{colortbl}
\usepackage{cleveref}
\usepackage{enumerate}
\usepackage{float}
\usepackage{forest}
\usepackage[htt]{hyphenat}
\usepackage{makecell}
\usepackage{multirow}
\usepackage{textgreek}
\usepackage{soul}
\usepackage{pifont}
\usepackage{subcaption} %
\usepackage{url}
\usepackage{xspace}
\usepackage{wrapfig}
\usepackage{paracol}
\usepackage{tabularx}
\usepackage{misc/picins}

\usepackage{tikz} %
\usetikzlibrary{automata} %
\usetikzlibrary{positioning} %
\usetikzlibrary{arrows} %
\usetikzlibrary{calc} %
\usetikzlibrary{patterns} %
\usetikzlibrary{snakes} %
\usetikzlibrary{shapes} %
\usetikzlibrary{arrows.meta} %

\newcommand{\code}[1]{\texttt{{\small #1}}}

\newcommand{\ie}{\textit{i.e.,\ }}
\newcommand{\eg}{\textit{e.g.,\ }}
\newcommand{\etal}{\textit{et al.\ }}
\newcommand{\etals}{\textit{et al.}'s\ }
\newcommand{\adhoc}{\textit{ad hoc}\xspace}

\newcommand{\vs}{{\em vs.}\xspace}

\usepackage{mdframed}
\mdfsetup{skipabove=0.5\topskip,skipbelow=0.5\topskip,align=center}

\usepackage{amsthm}
\newtheorem{thm}{Theorem}%

\usepackage{enumitem}
\setlist[itemize]{leftmargin=*,noitemsep,topsep=0pt}
\setlist[enumerate]{leftmargin=*}

\newcommand*\circled[1]{\tikz[baseline=(char.base)]{
            \node[shape=circle,draw,inner sep=0.5pt] (char) {#1};}}

\newlist{researchquestions}{enumerate}{1}
\setlist[researchquestions]{label*=\textbf{RQ\arabic*}, leftmargin=*}

\crefformat{section}{\S#2#1#3}
\crefmultiformat{section}{\S#2#1#3}{, \S#2#1#3}{, and \S#2#1#3}{}
\crefname{figure}{Figure}{Figures}
\crefname{appendix}{Appendix}{Appendices}
\crefname{table}{Table}{Tables}
\crefname{algorithm}{Algorithm}{Algorithms}
\crefname{listing}{Listing}{Listings}
\crefname{theorem}{Theorem}{Theorems}
\crefname{thm}{Theorem}{Theorems}
\crefname{lemma}{Lemma}{Lemmata}
\crefname{equation}{Eqt.}{Eqts.}
\crefformat{Grammar}{Grammar #1}

\ifDEBUG
  \PassOptionsToPackage{draft}{commenting}
\else
  \PassOptionsToPackage{final}{commenting}
\fi

\usepackage[nompar]{misc/commenting}

\declareauthor{EB}{Ethan}{green}
\declareauthor{CS}{Charlie}{magenta}
\declareauthor{JD}{Jamie}{blue}
\declareauthor{SC}{Sophie}{pink}
\declareauthor{BC}{Berk}{orange}
\declareauthor{DY}{Dongyoon}{cyan}

\authorcommand{EB}{comment}
\authorcommand{CS}{comment}
\authorcommand{JD}{comment}
\authorcommand{SC}{comment}
\authorcommand{BC}{comment}
\authorcommand{DY}{comment}

\onlyauthors{BC,CS,JD,EB,SC,DY} %

\newcommand{\toolname}{{\texttt{reuse-by-example}}\xspace}

\newcommand{\db}{{\textsc{RegexReuseDB}}\xspace}
\newcommand{\bench}{{\textsc{RegexCompBench}}\xspace}

\usepackage{xpatch}
\xpatchbibmacro{date}
  {\printtext[parens]{\printdate}}
  {\iffieldundef{date}{}{%
     \printtext[parens]{\printdate}}}
  {}{\typeout{Failed to patch date macro.}}

\usepackage{tcolorbox}

\begin{document}

\title[Is Reuse All You Need for Regex Composition?]{Is Reuse All You Need? A Systematic Comparison of Regular Expression Composition Strategies}

\author{Berk Çakar}
\orcid{0009-0006-6613-5591}
\affiliation{%
  \institution{Electrical and Computer Engineering, Purdue University}
  \city{West Lafayette}
  \state{IN}
  \country{USA}}
\email{bcakar@purdue.edu}

\author{Charles M. Sale}
\orcid{}
\affiliation{%
  \institution{Computer Science, Purdue University}
  \city{West Lafayette}
  \state{IN}
  \country{USA}
}
\email{cmsale@purdue.edu}

\author{Sophie Chen}
\orcid{0009-0000-4133-4910}
\affiliation{%
  \institution{Information Security, Carnegie Mellon University}
  \city{Pittsburgh}
  \state{PA}
  \country{USA}
}
\email{scchen@andrew.cmu.edu}

\author{Dongyoon Lee}
\orcid{0000-0002-2240-3316}
\affiliation{%
  \institution{Computer Science, Stony Brook University}
  \city{Stony Brook}
  \state{NY}
  \country{USA}
}
\email{dongyoon@cs.stonybrook.edu}

\author{James C. Davis}
\orcid{0000-0003-2495-686X}
\affiliation{%
  \institution{Electrical and Computer Engineering, Purdue University}
  \city{West Lafayette}
  \state{IN}
  \country{USA}
}
\email{davisjam@purdue.edu}

\authorsaddresses{%
  Authors' Contact Information: \href{https://orcid.org/0009-0006-6613-5591}{Berk Çakar}, bcakar@purdue.edu;
  \href{https://orcid.org/}{Charles M. Sale}, cmsale@purdue.edu;
  \href{https://orcid.org/0009-0000-4133-4910}{Sophie Chen}, scchen@andrew.cmu.edu;
  \href{https://orcid.org/0000-0002-2240-3316}{Dongyoon Lee}, dongyoon@cs.stonybrook.edu;
  \href{https://orcid.org/0000-0003-2495-686X}{James C. Davis}, davisjam@purdue.edu.%
}

\renewcommand{\shortauthors}{Çakar, Sale, Chen, Lee, and Davis}

\begin{abstract}
    Composing regular expressions (regexes) is a common but challenging engineering activity. Software engineers struggle with regex complexity, leading to defects, performance issues, and security vulnerabilities.
    Researchers have proposed tools to synthesize regexes automatically, and recent advances in large language models (LLMs) have also shown promise in generating regexes.
    Meanwhile, developers commonly reuse existing regexes from codebases and internet sources.
    No work to date has compared these various regex composition strategies, leaving software engineers unaware about which to use and researchers uncertain about open problems.

    We address this knowledge gap through a systematic evaluation of state-of-the-art regex reuse, formal synthesis, and LLM-based generation strategies.
    We curate a novel dataset of 901,516 regexes mined from open-source software projects and internet sources (\db), accompanied by a collection of 55,448 regex composition tasks defined by a target regex and its corresponding positive and negative string pairs (\bench).
    To address the absence of an automated regex reuse formulation, we design and implement \toolname, the first programming by example approach that leverages \db.
    Our evaluation then benchmarks \toolname, formal synthesizers, and LLMs on many aspects of interest to software engineers, including accuracy, maintainability, computational efficiency, and result diversity.
    Although all three approaches solve most composition tasks accurately, only \toolname and LLMs excel over the range of metrics we applied,
    and
    \toolname in particular offers engineers the variance in candidates that they say they find helpful.
    \textit{Ceteris paribus}, prefer the cheaper solution---for regex composition, perhaps reuse is all you need.
    Our findings provide actionable insights for developers selecting regex composition strategies and inform the design of future tools to improve regex reliability in software systems.
\end{abstract}


\maketitle

\section{Introduction}

Regular expressions (regexes) are widely used by software engineers to handle various string-matching tasks in software systems, as they are prevalent in many programming languages and development environments~\cite{singerExaminationSoftwareEngineering1997,michaelRegexesAreHard2019}.
Studies suggest that regexes appear in $\sim$30--50\% of software projects~\cite{chapmanExploringRegularExpression2016, davisImpactRegularExpression2018a, davis2019testing}.
Regex use cases range from web application firewalls \cite{ModSecurity,OWASPCoreRuleSet, WafRegexPatternSet} to input sanitization \cite{chapmanExploringRegularExpression2016,barlasExploitingInputSanitization2022,wangEmpiricalStudyRegular2020,eghbaliNoStringsAttached2020} and packet filtering \cite{roeschSnortLightweightIntrusion1999,xuSurveyRegularExpression2016, HowWeUse2022, PacketFilteringKeyword2022}.
Yet, despite their usefulness and prevalence, regexes are considered challenging to work with due to their complex syntax and semantics, which often lead to semantic errors, performance issues, and security vulnerabilities~\cite{wangEmpiricalStudyRegular2020,eghbaliNoStringsAttached2020,davisImpactRegularExpression2018a}.
Though composing them is difficult, engineers must use safe, correct regexes in their projects \cite{wangExploringRegularExpression2019}.
Poorly written regexes can expose systems to risks~\cite{goyvaerts2003runaway,roichman2009redos} particularly in web services~\cite{barlasExploitingInputSanitization2022,davisImpactRegularExpression2018a} and other regex-dependent applications.

To reduce human error in writing regexes from scratch, engineers pursue three strategies to develop regexes.
Some engineers reuse and adapt existing regexes based on their manual review of software repositories, forums, and curated libraries~\cite{baiExploringToolsStrategies2019,michaelRegexesAreHard2019,davisWhyArenRegular2019b}.
Others use formal synthesis tools that automatically generate regexes from input-output examples or natural language specifications (\eg~\cite{ferreiraFORESTInteractiveMultitree2021, bartoliCorrectionInferenceRegular2016, leeSynthesizingRegularExpressions2016, bartoliAutomaticSynthesisRegular2014, zhangInfeREStepbyStepRegex2023, parkSoftRegexGeneratingRegex2019, zhongSemRegexSemanticsBasedApproach2018, locascioNeuralGenerationRegular2016, Kushman2013UsingSU}).
\ifARXIV
\BC{Update the final list after updating the synthesizers in the background}
\fi
Most recently, advances in generative AI---particularly large language models (LLMs)---have enabled automated regex composition~\cite{siddiqUnderstandingRegularExpression2024, siddiqReGExDoS2024, tang2024enhancing}.
Despite the promise of these approaches, the literature lacks a systematic evaluation of their relative merits for real-world regex composition tasks.
This gap leaves practitioners without guidance and researchers without clear objectives for improvement.

To address this gap, we make three contributions.
First, we curate new resources:
  (i) \db, a large-scale database of production-ready regexes,
  and
  (ii) \bench, a benchmark of real-world regex composition tasks with associated input-output examples.
Second, we introduce \toolname, the first system for automated regex reuse, operationalized as a programming by example approach~\cite{liebermanProgrammingExampleIntroduction2000,liebermanYourWishMy2001} that leverages \db. %
Third, with these in hand, we systematically compare reuse, formal synthesis, and LLM-based generation strategies on \bench across multiple metrics relevant to software engineering, including accuracy, maintainability, diversity, and computational efficiency.
This provides a comparative evaluation of regex composition strategies with actionable insights for both practitioners and researchers.

Our evaluation indicates that for the regular expression composition tasks used in typical open-source software, reuse may be a more effective technique than LLMs.
Reuse achieves the highest overall effectiveness: it matches the (near-perfect) functional accuracy of state-of-the-art LLMs while producing a markedly broader range of candidate regexes.
This variance enables engineers to trade off non-functional properties such as pattern length, feature usage, and automaton complexity according to preferences project requirements.
In contrast, formal synthesis tools could only solve $\sim$80\% of the tasks.

Our primary contributions are as follows:

\begin{enumerate}
    \item We share two new resources: \db, a curated database of 900K+ regexes mined from open-source software, Stack Overflow, and RegExLib; and \bench, the first real-world benchmark of 55K+ regex composition tasks with associated input-output examples.
    \item We introduce \toolname, the first system for the automated regex reuse. %
    \item Supported by a user study, we emphasize \textit{candidate set variance} as the primary means of interpreting the results of regex composition methods on non-functional properties.
    \item We perform the first comprehensive comparison of regex reuse, synthesis, and generation strategies.
      Our results show that reuse and LLM-based generation achieve high accuracy, but differ in maintainability and robustness, informing future tool design and engineering practice.
\end{enumerate}

\noindent
\textbf{\ul{Significance}}: Our study is the first to systematically evaluate different regex composition strategies against real-world tasks.
  At present, \toolname and state-of-the-art LLMs achieve comparable (near-perfect) accuracy on \bench, but the reuse approach better satisfies engineers' preferences related to efficiency and diversity of regex solution candidates.
We therefore recommend that engineers adopt reuse-based approaches for mundane regex composition tasks, and use LLM-based approaches for unusual use-cases lacking existing solutions.
For researchers, our findings suggest that formal synthesis algorithms are now outdated, and that future regex composition tools might combine the benefits of real-world regexes with the automated exploration capabilities of LLMs.
Our \db and \bench resources will enable these and other studies on regex engineering.

\section{Background and Related Work} \label{sec:background}
We
  discuss regexes in software engineering (\cref{subsec:regular-expressions}),
  review regex composition strategies (\cref{subsec:regex-composition-strategies}),
  and
  summarize past benchmarking of regex composition (\cref{subsec:regex-composition-benchmarks}).

\subsection{Regular Expressions} \label{subsec:regular-expressions}

\subsubsection{K- and E-regexes}
Regexes describe languages: sets of strings, drawn from an underlying alphabet, that conform to specified patterns~\cite{hopcroftIntroductionAutomataTheory2007}.
Programming languages and tools support either:

\begin{itemize}
\item \textbf{Kleene's original regexes (\textit{K-regexes})}~\cite{Kleene1951NerveNetsAndRegularLanguages},
  in which a regex pattern $R$ semantically represents the language $L(R)$ through the following defining operations: concatenation ($R_1R_2$), disjunction ($R_1\cup{R}_2$) and repetition ($R*$).
  These constructs have led to a rich theory of regexes~\cite{Hopcroft2006}.
  Many syntactic sugars for K-regexes exist, \eg custom character classes~\cite{friedlMasteringRegularExpressions2002}.
  For example, the regex semantics of Rust and Go support only K-regexes.
\item \textbf{Extended regexes (\textit{E-regexes})}~\cite{friedlMasteringRegularExpressions2002,Hazel1997PCRE,Aho1980RegexChapter}
   introduce additional operators, notably look-around assertions (\eg ``(?=$R$)'') and backreferences (\eg ``$(R)\ldots\setminus{}$1'').
  These features improve expressiveness, at the cost of computational efficiency and theoretical tractability.
  The regex semantics of most programming languages (\eg JavaScript, Java, Python, and C\#) support E-regexes.
\end{itemize}

\noindent
K-regexes can be represented by both nondeterministic and deterministic finite automata (NFA and DFA, respectively)~\cite{mcnaughton_regular_1960,thompson_programming_1968,cox2007Blog}.
Applying classical automata theory to E-regexes is challenging because some E-regex features exceed the expressive power of regular languages~\cite{Aho1980RegexChapter}, though not all do so~\cite{chattopadhyay_verified_2025}.
Regex engines that implement E-regex matching instead rely on techniques such as recursive descent parsing~\cite{davis2021using}.

\subsubsection{Regular Expressions in Software Engineering} \label{subsubsec:regular-expressions-in-software-engineering}

Software engineers employ regexes for text processing (\eg searching, matching, extracting, and manipulating strings).
A common application is in
 input sanitization~\cite{chapmanExploringRegularExpression2016,barlasExploitingInputSanitization2022,wangEmpiricalStudyRegular2020,eghbaliNoStringsAttached2020}
  and validation (\eg when defining OpenAPI input specifications~\cite{barlasExploitingInputSanitization2022}) and packet filtering~\cite{roeschSnortLightweightIntrusion1999,xuSurveyRegularExpression2016, HowWeUse2022, PacketFilteringKeyword2022}.
Regexes are also common in machine learning workflows and data cleaning pipelines, where they enable automated extraction, normalization, and validation of textual features from raw datasets~\cite{tf_regex_full_match,pyspark_regexp}.
Engineers also use regexes in
  web application firewalls such as ModSecurity~\cite{ModSecurity,OWASPCoreRuleSet} and AWS WAF~\cite{WafRegexPatternSet},
  intrusion detection and prevention systems like Snort~\cite{roeschSnortLightweightIntrusion1999, RegexSnortRule},
  antivirus software including ClamAV~\cite{LogicalSignaturesClamAV},
  and
  log auditing tools (\eg New Relic~\cite{buchananRegexParsingExtract2023}, Coralogix~\cite{razRegEx101Guide2019}, Logstash~\cite{AccessingEventData, GrokFilterPlugin}).
Reflecting their broad use cases, regexes are reportedly used in 30--50\% of Python, Java, and JavaScript projects~\cite{chapmanExploringRegularExpression2016, davisImpactRegularExpression2018a, davis2019testing}.

When composing regexes, software engineers sometimes write them from scratch.
However, because many engineers find regexes hard to read, validate, and document~\cite{michaelRegexesAreHard2019}, they commonly reuse regexes or apply automated composition techniques (\cref{subsubsec:reusing-regexes}). %
Next, engineers iteratively test and refine the behavior of candidate regexes using visualization and debugging tools such as RegExper~\cite{RegExper}, Regex101~\cite{Regex101BuildTest}, and RegViz~\cite{beckRegVizVisualDebugging2014, RegVizVisualDebugging}.
To validate the resulting regexes, engineers identify test inputs and edge cases using input generation tools~\cite{veanes2010rex,larsonGeneratingEvilTest2016,arcaini2017mutrex} and coverage measures~\cite{wangHowWellAre2018}. %

\subsubsection{Common Challenges} \label{subsubsec:common-challenges}

Engineering challenges appear at all stages of the regex development process.
In problem definition, software engineers often struggle to balance between overly permissive patterns that accept unintended inputs and overly restrictive patterns that reject legitimate ones~\cite{michaelRegexesAreHard2019}.
As developers implement regexes, they are often confused by inconsistent regex syntax and semantics across different programming environments~\cite{chapmanExploringRegexComprehension2017,davisWhyArenRegular2019b}.
During validation, traditional testing tools do not include coverage metrics that capture the automaton-level behavior of regexes, resulting in undertesting when developers rely solely on source-level line or branch coverage~\cite{wangHowWellAre2018}.
Incorrect regex semantics (\ie being too permissive or too restrictive) and inadequate testing are common causes of string-related defects~\cite{wang_demystifying_2022,wangEmpiricalStudyRegular2020,eghbaliNoStringsAttached2020}.

Using the Common Weakness Enumeration (CWE) taxonomy for software weaknesses, regex defects create two kinds of weaknesses in the software systems in which they are used:
\begin{itemize}
    \item \textbf{CWE-20}:
      Regex flaws in input validation can result in CWE-20: \textit{Improper Input Validation}~\cite{mitre_cwe20}. %
    If the pattern is too permissive, malicious strings can bypass checks; too restrictive and legitimate user input is blocked.
    \item \textbf{CWE-1333}:
      Some regexes have super-linear match performance in regex engines that use a backtracking search~\cite{staicuFreezingWebStudy2018,davisImpactRegularExpression2018a}.
      On malicious inputs, this property can cause Regular Expression Denial of Service (ReDoS)~\cite{bhuiyan_sok_2025}, \ie CWE-1333: \textit{Inefficient Regular Expression Complexity}~\cite{mitre_cwe1333}.
\end{itemize}

\subsection{Regex Composition Strategies} \label{subsec:regex-composition-strategies}

Apart from writing regexes from scratch, engineers rely on three main regex composition strategies:
  reuse (adapting existing regexes),
  formal synthesis (algorithmic construction),
  and
  LLMs (learned probabilistic generation).
Together, these approaches represent the spectrum of current practices.

\begin{tcolorbox}[colback=gray!5,colframe=black,title=Definition 1: Regular Expression Composition]
\label{def:RegexComposition}
Regular expression composition indicates a string-matching task specified by a set of positive examples $P$ and negative examples $N$.
Here, the goal is to produce one or more candidate regexes $R$ that match all strings in $P$ while rejecting those in $N$:
\[
    compose(P, N) \rightarrow R.
\]
Compared to prior work, this definition omits other possible inputs (such as natural language- or automaton-based specifications), which user studies suggest are unwieldy for typical software engineers~\cite{michaelRegexesAreHard2019,chapmanExploringRegexComprehension2017}.
However, this definition acknowledges the inherent gap between example-based specifications and the engineer's actual intent, as noted by Michael \etal~\cite{michaelRegexesAreHard2019}.
\end{tcolorbox}

\subsubsection{Regex Reuse} \label{subsubsec:reusing-regexes}
To address common tasks (\eg matching emails), developers often reuse and adapt existing regexes, drawing from internet sources such as Stack Overflow~\cite{NewestRegexQuestions} and RegExLib~\cite{RegularExpressionLibrary}, as well as from their own codebases or other open-source projects.
Engineers seek trusted sources for higher quality and better-tested regexes~\cite{davisWhyArenRegular2019b, michaelRegexesAreHard2019}.
In a 2019 survey~\cite{davisWhyArenRegular2019b}, 94\% of developers reported reusing regular expressions, with 50\% indicating they reuse regexes at least half the time. %

Yet, it is often difficult to find appropriate regexes for reuse.
Although developers often begin with searches in natural language, most regex composition tasks are difficult to articulate in search terms~\cite{michaelRegexesAreHard2019}.
Platforms and tools like RegExLib and \textsc{Resem}~\cite{takeshigeResemSearchingRegular2022} try to help developers find the relevant regexes by providing tags (\eg ``email'', ``phone number''), descriptions, and input-output examples.
When developers find potentially reusable regexes, they usually prefer simpler patterns that are easier to understand and adapt~\cite{michaelRegexesAreHard2019}.
No example-oriented regex reuse framework has yet been developed, which hinders the evaluation of its potential.

\subsubsection{Formal Regex Synthesis} \label{subsubsec:regex-synthesis}
Regex synthesis is a longstanding algorithmic approach to automatically generating regexes by leveraging specifications such as natural language descriptions or example strings. %
Multi-modal synthesis frameworks like SemRegex~\cite{zhongSemRegexSemanticsBasedApproach2018}, \textsc{Regel}~\cite{chenMultimodalSynthesisRegular2020}, \textsc{DeepSketch}~\cite{yeSketchDrivenRegularExpression2020} \textsc{TransRegex}~\cite{li2021transregex}, and \textsc{Smore}~\cite{chenDataExtractionSemantic2023} combine both specifications to overcome the inherent ambiguity of either method alone, utilizing semantic understanding rather than mere syntactic pattern matching.
These systems employ neural networks or SMT solvers to navigate the complex search space of possible regex solutions that satisfy the constraints given.

Regex synthesis approaches can be further categorized based on their primary objectives and interaction models.
For example, repair-oriented synthesizers like RFixer~\cite{panAutomaticRepairRegular2019} focus on refining existing regexes to accommodate new examples, while \textsc{AlphaRegex}~\cite{leeSynthesizingRegularExpressions2016} uses symbolic search techniques to guarantee correctness.
Because formal synthesis methods struggle with generalization,
interactive synthesizers such as \textsc{Forest}~\cite{ferreiraFORESTInteractiveMultitree2021} incorporate human feedback loops to iteratively refine the generated regexes.
Despite myriad techniques, there is no evidence that developers actually use regex synthesis to compose regexes~\cite{chapmanExploringRegularExpression2016,michaelRegexesAreHard2019}.

\subsubsection{LLMs for Regex Generation} \label{subsubsec:generative-ai-for-regexes}
LLMs have emerged as powerful tools for regex generation~\cite{siddiqReGExDoS2024,siddiqUnderstandingRegularExpression2024,tang2024enhancing}. %
Unlike traditional regex synthesis methods that rely on precise specifications, LLMs can interpret unstructured examples and natural language descriptions to generate the corresponding regex patterns, which makes regex composition easier for developers.
LLMs may be considered as a form of automated reuse, since they have been trained on publicly available code, discussions, and documentation~\cite{touvron_llama_2023,hui_qwen25-coder_2024}, each containing regexes. %

Recent developer discussions suggest that LLMs are being employed for regex composition tasks, much as they are for other programming activities~\cite{houLargeLanguageModels2024,fanLargeLanguageModels2023, stack_1, stack_2, stack_3}.
However, the extent of real-world adoption and the effectiveness of LLM-based regex composition remain unknown.

\subsection{Benchmarking of Regex Composition}
\label{subsec:regex-composition-benchmarks}

The evaluation of different regex composition strategies remains fragmented and limited.
Studies of formal synthesis and, more recently, LLMs have consistently relied on small-scale or synthetic benchmarks.
For instance, synthesis studies use small benchmarks of
  75 real-world tasks~\cite{panAutomaticRepairRegular2019},
  $\sim$1,000 real-world tasks~\cite{ye2020benchmarking};
  or
  synthetic tasks~\cite{kim2021splitregex};
  or a mix~\cite{li2021transregex}.
Similarly, only two works for LLMs~\cite{siddiqReGExDoS2024,siddiqUnderstandingRegularExpression2024} showed small-scale demonstrations with limited benchmarks (1,001 composition tasks from RegExLib).
Most of these works focus on the functional property of accuracy, \ie whether the regex composition method satisfies the specified task.
Some also utilize automaton-focused metrics such as ANMLZoo~\cite{wadden2016anmlzoo}, rather than metrics interpreted for software quality and maintainability.
Meanwhile, regex reuse has only been described qualitatively~\cite{michaelRegexesAreHard2019,davisWhyArenRegular2019b}, with no measurement study of its effectiveness at all.

What has not been done is a systematic comparison of all three approaches on a large real-world dataset.
In addition, studies rarely use software engineering-relevant regex metrics such as those proposed by
  Chapman, Wang, and Stolee~\cite{chapmanExploringRegularExpression2016,chapmanExploringRegexComprehension2017,wangHowWellAre2018}
  and
  Davis \etal~\cite{davis2019testing,michaelRegexesAreHard2019}, which require non-automata approaches or alternative interpretations of automata metrics (\cref{subsec:regex-metrics}).

\section{Research Questions and Study Overview}
A critical knowledge gap exists in the literature on different regex composition strategies: there is no systematic comparison between reuse, formal synthesis, and LLM-based approaches.
This lack of comparative evaluation
  leaves engineers without guidance in selecting a regex composition strategy
  and
  prevents researchers from identifying objectives for future work.
Engineers are concerned with both
  the \textit{functional properties} of the resulting regex (\eg whether it solves the problem at hand)
  and
  its \textit{non-functional properties} (\eg whether it is overly complex).
We therefore ask:

\begin{researchquestions}
    \item How do different regex composition strategies compare in terms of \textit{functional properties}?
    \item How do different regex composition strategies compare in terms of \textit{non-functional properties}?
\end{researchquestions}

To answer these questions, we
  create a benchmark dataset (\cref{sec:data-collection}),
  operationalize programming by example for regex reuse (\cref{sec:reuse-by-example}),
  curate metrics and baselines (\cref{sec:Methodology}),
  and
  run the measurements (\cref{sec:results}).

\section{Data Collection: \db and \bench} \label{sec:data-collection}

\begin{figure}
    \centering
    \includegraphics[trim=18 10 35 3, clip,width=1\linewidth]{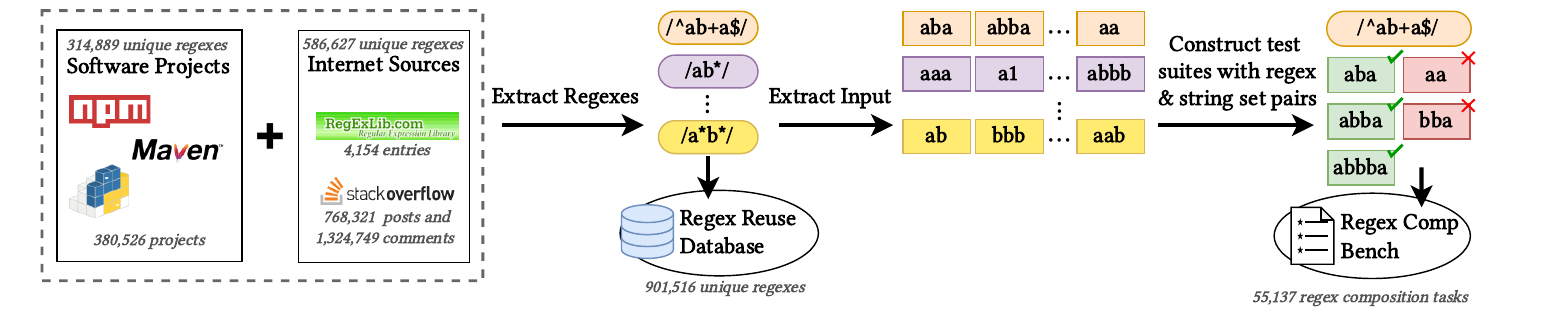}
    \caption{
    Data collection methodology to construct \db and \bench.
    }
    \label{fig:data-collection}
\end{figure}

In this section, we describe the curation of our two new resources: \db (\Cref{subsec:db}) and \bench (\Cref{subsec:bench}).
\Cref{fig:data-collection} illustrates our data collection process.

\subsection{\db} \label{subsec:db}

\subsubsection{Goal}
Our goal with \db is to provide a large-scale corpus of production-ready regexes for use in regex composition research.
The most similar datasets for this goal are those published by Davis \etal in 2018~\cite{davisImpactRegularExpression2018a} and 2019~\cite{davis2019testing}, which contain 300,000--500,000 regexes mined primarily from GitHub.
In a related work, Davis \etal~\cite{davisWhyArenRegular2019b} mined some regexes from Stack Overflow, but in a relatively \adhoc manner.

We contribute two improvements over those datasets.
First, it has been over 5 years since their publication, and we provide an updated snapshot of regexes from GitHub.
Second, we systematically mine Stack Overflow, addressing two shortcomings of the prior effort.

\subsubsection{Approach}

We mined regexes from open-source software projects across three programming language ecosystems (Java's Maven, Java\-Script's NPM, and Python's PyPI).
We also mine regexes from two common internet sources, RegExLib and Stack Overflow.

\paragraph{Mining Open-Source Software Projects}
We selected repositories from three of the largest programming language ecosystems: Java (Maven), JavaScript (NPM), and Python (PyPI).
We excluded packages with fewer than 100 downloads and then randomly sampled 50\% of the remaining packages from each ecosystem.
For each selected package, we performed static extraction to identify all regexes appearing in source code.
We followed Davis \etals methodology~\cite{davisWhyArenRegular2019b}, identifying regex constructors and extracting the static strings used to initialize them.

In total, we mined 4,783,558 regexes in 380,526 projects, of which 314,889 were unique.

\paragraph{Mining Internet Sources}
As a second source, we mined internet sources that developers commonly use to find and share regexes~\cite{davisWhyArenRegular2019b}. We focused on two primary sources: (i) RegExLib, a curated online library of regexes, and (ii) Stack Overflow, a widely-used question-and-answer forum for developers.
For RegExLib, we scraped all available regexes on the platform.
For Stack Overflow, we gathered regexes appearing in posts tagged with the \texttt{regex} keyword, as well as from their associated comments, by identifying regex syntax within HTML \texttt{<code>} elements.\footnote{A notable weakness of the prior approach~\cite{davisWhyArenRegular2019b} was the risk of false positives: \texttt{<code>} elements contain all kinds of code snippets, not just regexes. Identifying the regex within using heuristics was difficult in 2019. We address this with an LLM-based pipeline that extracts regexes from such snippets.}
The previous work did not mine comments, which was a significant oversight --- the comments on Stack Overflow often provide competing solutions, the very type of data that could benefit automated regex composition.

From RegExLib we examined 4,154 entries comprising 3,878 unique regexes.
From Stack Overflow we examined 768,321 posts, which included 882,220 regexes --- 472,951 of these were unique and not already in open-source software projects and RegExLib.
Our subsequent processing of 1,324,749 associated comments yielded 294,519 regexes, adding a further 109,798 unique regexes.

\subsubsection{Summary}

\Cref{tab:db-stats} summarizes \db: a corpus of over 900K unique regexes.

\parpic[r]{%
  \begin{minipage}{0.45\textwidth}
    \centering
    \footnotesize
    \captionof{table}{Breakdown of the \db corpus. For each source, we list the number of targets mined (\eg projects or posts), the total regexes found, and the number of unique regexes contributed to the database.}
    \label{tab:db-stats}
    \vspace{-2mm}
    \begin{tabular}{@{}l r r r@{}}
      \toprule
      \textbf{Source} & \textbf{Targets} & \textbf{Regexes} & \textbf{Contrib.} \\
      \midrule
      OSS Projects & 380,526 & 1,307,691 & 314,889 \\
      \midrule
      RegExLib     & 4,154   & 4,154     & 3,878   \\
      SO Posts     & 768,321 & 882,220   & 472,951 \\
      SO Comments  & 1,324,749 & 294,519 & 109,798 \\
      \midrule
      \textbf{Total} & \textbf{2,477,750} & \textbf{2,488,584} & \textbf{901,516} \\
      \bottomrule
    \end{tabular}
    \vspace{-10mm}
  \end{minipage}
}

\subsection{\bench} \label{subsec:bench}

\subsubsection{Goal}
The burden of this paper is to systematically compare the three competing approaches to regex composition.
To do this, we need a large-scale benchmark of real-world regex composition tasks.
With reference to \hyperref[def:RegexComposition]{Definition~1}, we need sets of positive and negative examples and the accompanying ground truth --- the regexes that engineers developed based on those examples.
Existing benchmarks are small or synthetic and do not provide a useful starting point (\cref{subsec:regex-composition-benchmarks}).

\subsubsection{Approach}

We create \bench by modeling regex composition tasks via \textit{regex testing practices}.
Why is this suitable?
Prior empirical studies have found that regexes are often developed as the engineer iteratively refines their understanding of the input domain~\cite{michaelRegexesAreHard2019,eghbaliNoStringsAttached2020,wang_demystifying_2022}.
The test cases serve as the bounds on the desired behavior, \ie a living specification~\cite{wangHowWellAre2018,wangExploringRegularExpression2019}.

\paragraph{Test Suite Extraction from Open-Source Software Projects} For software packages, we employed dynamic analysis to capture regex test cases.
Imitating the approaches of
  Wang \& Stolee~\cite{wangHowWellAre2018}
  and
  Davis \etal~\cite{davis2019testing}, we instrumented the runtimes of the programming languages to log the strings passed to each regex while executing the each project's test suite.
We used standard test suite execution commands for each language ecosystem (\eg \code{npm run test} for NPM).
This process yielded 51,414 regex composition tasks from open-source software projects.

\paragraph{Test Suite Extraction from Internet Sources} For internet sources, we only considered RegExLib.
While Stack Overflow is a valuable source of regexes, the examples are embedded in unstructured natural language, making them difficult to parse reliably for a benchmark.
In contrast, RegExLib provides structured lists of positive and negative examples alongside its regexes.
Therefore, our earlier mining of 4,154 regexes has already provided us with corresponding test suites.
This contributed an additional 4,034 regex composition tasks, completing our benchmark of 55,448 regex composition tasks.

\subsubsection{Summary}

\bench is the first large-scale real-world benchmark of regex composition tasks.
It comprises 55,448 regex composition tasks, each defined by a ground-truth regex and a corresponding test suite of positive (matching) and negative (non-matching) example strings that specify its intended behavior.
The median task has 6 example strings in total.

\section{\toolname: The First Automated Approach for Regex Reuse} \label{sec:reuse-by-example}

\begin{figure}
    \centering
    \begin{subfigure}[t]{0.475\textwidth}
        \centering
        \includegraphics[width=\linewidth]{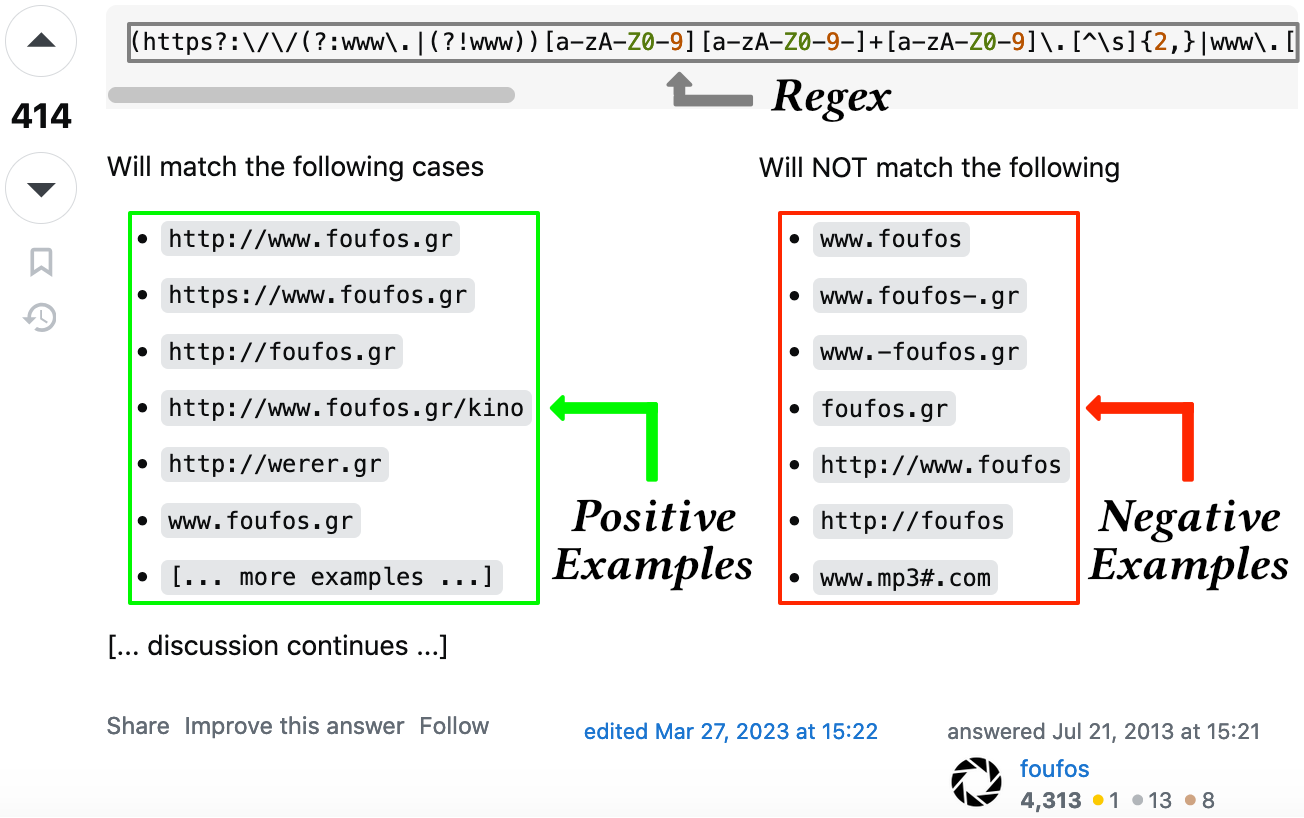}
        \caption{
        Example of regex reuse from Stack Overflow.
        Engineers often accompany regexes with example strings to illustrate intended behavior (programming by example).
        }
        \label{fig:regex-reuse-examples}
    \end{subfigure}
    \hfill
    \begin{subfigure}[t]{0.475\textwidth}
        \centering
        \includegraphics[width=\linewidth]{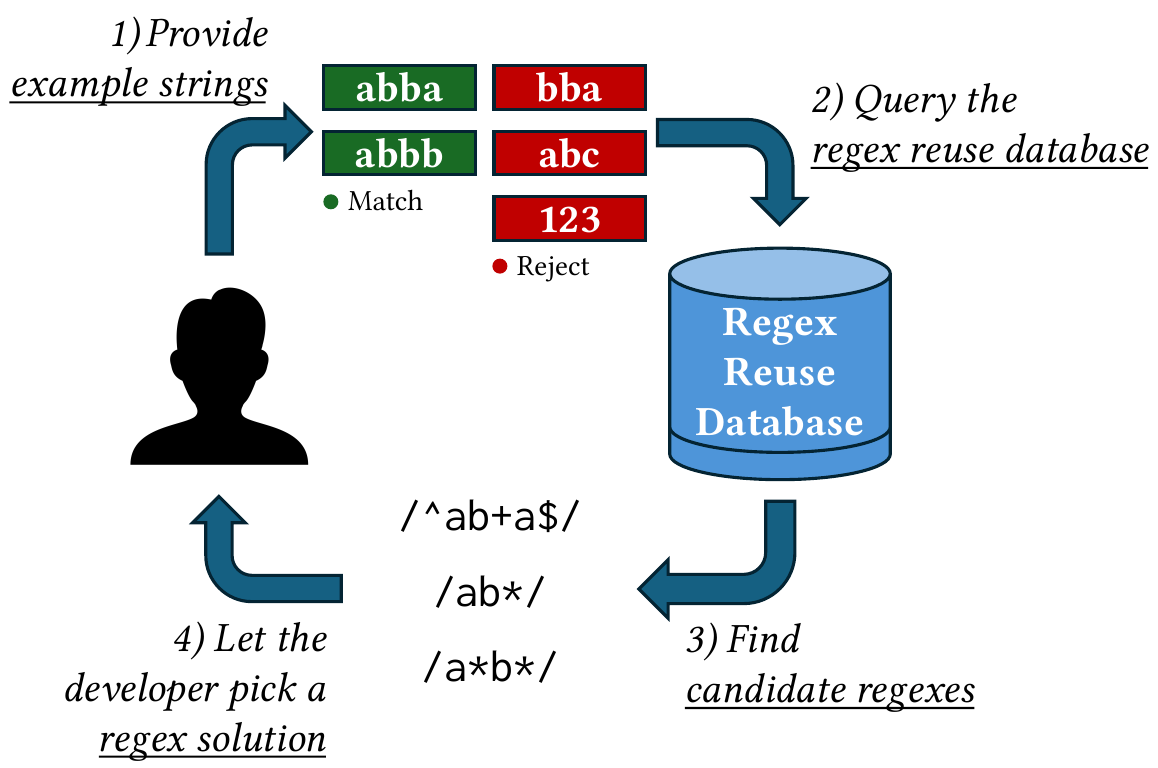}
        \caption{
        The \toolname workflow.
        Users specify examples, the system retrieves matching regex candidates, and developers select or refine their search.
        }
        \label{fig:reuse-by-example-workflow}
    \end{subfigure}
    \vspace{-0.1cm}
    \caption{
    Illustration of
    (a) real-world regex reuse,
    and
    (b) our \toolname workflow.
    }
    \label{fig:reuse-double-figure}
  \end{figure}

We propose \toolname as a concrete embodiment of the regex reuse concept (\cref{subsubsec:reusing-regexes}), permitting a comparison with existing automated regex composition methods. %
\Cref{subsec:requirements} discusses how \toolname fulfills key requirements drawn from the regex engineering practice and \Cref{subsec:operationalization} overviews how \toolname operationalizes regex reuse.
\cref{fig:reuse-double-figure} illustrates our approach.

\subsection{Requirements}
\label{subsec:requirements}

\toolname addresses two requirements from regex engineering practice~\cite{michaelRegexesAreHard2019}.
\ul{First}, engineers need a semantically precise way to search for regexes, as natural language searches are inadequate for describing complex pattern-matching scenarios.
\ul{Second}, engineers require not just any functional solution, but one that offer appropriate strictness.
As one engineer told Michael \etal~\cite{michaelRegexesAreHard2019}: \textit{``What might be tricky is deciding whether or not you want to match too much or too little''}.
\cref{fig:regex-reuse-examples} shows a typical example of how engineers seek and describe regexes.

\subsection{Operationalization}
\label{subsec:operationalization}

To address the first requirement, we take a programming by example approach.
Programming by example is a program synthesis approach in which developers provide semantic input-output examples to automatically generate code that satisfies their requirements~\cite{liebermanProgrammingExampleIntroduction2000, liebermanYourWishMy2001}.
This technique bridges the gap between what engineers can easily express (concrete examples) and what they need (executable code)~\cite{michaelRegexesAreHard2019,davisWhyArenRegular2019b}.
In the context of regexes, as described in~\cref{def:RegexComposition}, the input-output examples are
  strings that should be matched (positive) or rejected (negative) by the desired regex.
In this framing, the second requirement, of choosing the proper regex for the task, becomes the task of choosing among the candidates that satisfy the query.

Our version of \toolname uses a database of engineer-developed regexes mined software projects and Internet sources (\db).
\cref{fig:reuse-by-example-workflow} shows the main components:

\vspace{0.5em}
\begin{itemize}

    \item \textbf{Query Language (Step \circled{1})}: Users provide positive and negative example strings that define their regex composition task. These examples project the user's actual task into the I/O space. %
    \item \textbf{Search Process (Steps \circled{2}--\circled{3})}: \toolname matches the example constraints against the regex reuse database, identifying potential candidate regexes that satisfy the specified conditions.
    \item \textbf{Candidate Selection (Step \circled{4})}:
      To assist engineers in selection, our version of \toolname ranks the candidates by strictness (\ie the amount of the candidate's automaton covered by the I/O examples).
      This scheme can help users find candidates with varying degrees of constraint (broader or narrower matching behavior), aligning with engineers' second requirement. %
  \end{itemize}

\vspace{-0.2cm}
\section{Methodology} \label{sec:Methodology}

We describe
 our evaluation hardware (\cref{subsec:eval-hardware}),
 metrics for comparison (\cref{subsec:regex-metrics}),
 sampling method from \bench (\cref{subsec:experimental-data},
 and
 selection of representatives for each regex composition strategy (\cref{sec:ExperimentSetup-ComparisonTargets}).

\subsection{Hardware and Commercial LLM Access} \label{subsec:eval-hardware}

For the \toolname and formal synthesis approaches, we ran on a bare-metal Dell Precision 5820 Tower server, equipped with an Intel Xeon W-2295 CPU (18 cores, 36 threads) and 188GB of RAM.
For the LLM-based approaches, we accessed commercial services via OpenRouter~\cite{OpenRouter} and ran open-source models via vLLM~\cite{vllm-project/vllm_2025} on a cluster with 8 NVIDIA H100 GPUs.

\subsection{Metrics for Evaluating Regex Composition Strategies} \label{subsec:regex-metrics}

In this work, we assess regex composition strategies by their performance on real-world tasks.
We define such a \textit{regex composition task} as a regex written by an engineer, along with its test suite of positive (matching) and negative (non-matching) example strings that define the intended behavior.
This definition follows from \hyperref[def:RegexComposition]{Definition~1}.

The measurements discussed in this section were implemented using the Python regex engine for semantic properties and Python's regex grammar for syntactic analysis.
We discarded regex candidates that did not work in these environments, whether from \bench or the various composition strategies.
Some of these measures are defined in terms of a single regex, while others are comparative.
For comparative metrics, we use the engineer's solution as the reference. %
\Cref{tab:evaluation-metrics} summarizes the metrics used in this study.

\begin{table}
    \captionof{table}{
    Metrics used to evaluate regex composition strategies. For each measure, we list representative papers from the literature where the metric was used.
    \textit{\textbf{Notation:}} $R$ denotes the regex; \texttt{compose} is the composition function; $P$ and $N$ are the positive and negative examples \texttt{Aprx} is a language approximation function; $R_{\text{Ground}}$ is the ground truth regex developed by the engineer; \texttt{AST} builds $R$'s abstract syntax tree; \texttt{NFA} constructs the non-deterministic finite automaton of $R$; and \texttt{EditDist} is a syntax-based distance function.
    }
    \label{tab:evaluation-metrics}
    \vspace{-0.25\baselineskip}
    \centering
    \footnotesize
    \begin{tabularx}{\linewidth}{@{} p{1.2cm} p{1.75cm} p{2.75cm} p{5.4cm} c @{}}
      \toprule
      \textbf{Category} & \textbf{Metric} & \textbf{Description} & \textbf{Definition} & \textbf{Lit.} \\
      \midrule
      \multirow{3}{*}{\textbf{Functional}}
      & Accuracy & Satisfies examples $P$, $N$ &
      $\frac{|\{x \in P : R(x) = 1\}| + |\{x \in N : R(x) = 0\}|}{|P| + |N|}$
       &  \cite{siddiqUnderstandingRegularExpression2024,chenMultimodalSynthesisRegular2020,li2021transregex} \\
      & Semantic sim. & Overlap w/ground truth &
      {\footnotesize $\frac{|\{x \mid x \in \texttt{Aprx}(R) \cup \texttt{Aprx}(R_{\text{ground}}),\ R(x) = 1\ \wedge\ R_{\text{ground}}(x) = 1\}|}{|\texttt{Aprx}(R) \cup \texttt{Aprx}(R_{\text{ground}})|}$ }
       &  \cite{ parkSoftRegexGeneratingRegex2019,zhongSemRegexSemanticsBasedApproach2018,yeSketchDrivenRegularExpression2020} \\
      & Generation eff. & Time to compose &
       $\texttt{time}(\texttt{compose}(P, N))$
        &  \cite{li2021transregex,leeSynthesizingRegularExpressions2016} \\
      \midrule
      \multirow{4}{*}{\textbf{Non-Funct.}}
      & Pattern length & Length of the regex &
      $\texttt{length}(R)$
      &  \cite{davis2019testing} \\
      & Regex feats. & Distinct constructs used &
      $\left|\{\text{c} \mid \text{c is a construct used in \texttt{AST}}(R)\}\right|$
      &  \cite{davis2019testing} \\
      & Syntactic sim. & Edit dist. to ground truth &
      $\texttt{EditDist}(\texttt{AST}(R), \texttt{AST}(R_{\text{Ground}}))$
      &  \cite{siddiqUnderstandingRegularExpression2024,zhongSemRegexSemanticsBasedApproach2018} \\
      & Automaton size & Language complexity &
      $|\mathcal{V}| + |\mathcal{E}| \text{ of } \texttt{NFA}(R)$
      &  \cite{wangHowWellAre2018,davis2019testing} \\
      \bottomrule
    \end{tabularx}
    \vspace{-3mm}
  \end{table}

\subsubsection{Functional Metrics}

\leavevmode\par\vspace{0.1\baselineskip}

\noindent Functional metrics evaluate a candidate's behavior relative to the engineer's intent.

\ul{Accuracy:}
This is the direct measure of correctness, quantifying conformance to the task (accepting all positives and rejecting all negatives)~\cite{siddiqUnderstandingRegularExpression2024,chenMultimodalSynthesisRegular2020,li2021transregex}.
We measure a $[0,1]$ ratio.

\ul{Semantic similarity:}
The interpretation of an accuracy measure is limited by the comprehensiveness of the test suite in \bench.
To go beyond a fixed test suite, we measure semantic similarity, which compares the languages of the candidate and ground-truth regexes.
Prior works propose automata-based equivalence~\cite{parkSoftRegexGeneratingRegex2019,zhongSemRegexSemanticsBasedApproach2018,yeSketchDrivenRegularExpression2020} and language approximations~\cite{larsonGeneratingEvilTest2016}.
We believe language approximation is easier to interpret,
and we use a language approximation that generates automaton-covering strings and assumes symbolic edges.

\ul{Generation efficiency:}
Finally, engineers consider how long it takes to generate candidates.
We time how long each strategy takes to produce candidate regexes.
For strategies run on our machines, we report the wall clock time to compute on a dedicated core.
For strategies run on cloud services, we report the end-to-end latency.
For LLMs, we specifically report the latency until we obtain the first working candidate, which may include multiple rounds of feedback to the LLM.

\vspace{-0.1cm}
\subsubsection{Non-Functional Metrics}
\label{sec:metrics-nonFunctional}

Non-functional metrics capture qualities related to comprehension and maintainability.
We selected metrics from a variety of literature.
We focused on metrics with direct bearing on the software engineer's experience~\cite{davis2019testing}, and included other measures (\eg automaton size) to permit comparison to the regex engine literature~\cite{wadden2016anmlzoo}.

\ul{Pattern length} and \ul{Number of regex features} are measures of the comprehensibility of the regex~\cite{chapmanExploringRegexComprehension2017}.
Novices find longer patterns and those with more features to be difficult to understand, though more skilled engineers may disagree.
We measure these using the \code{length} function on the string representation of a regex, and through counting the distinct node types in the abstract syntax tree after the regex is parsed.

\ul{Syntactic similarity:}
This is a comparison between a candidate regex and the ground truth.
Specifically, we measure the edit distance between their respective parse trees.
We implemented this using the standard Zhang \& Shasha tree edit distance~\cite{zhangSimpleFastAlgorithms1989}.

\ul{Automaton size:}
This measures the formal complexity of the regex by its automaton representation.
Specifically, we measure the size of the NFA representation, defined as the sum of its vertices and edges.
Our implementation uses the Pyformlang~\cite{Romero_2021,pyformlang_github} library, which generates a non-symbolic automaton where character classes like \code{[a-z]} or \code{{\textbackslash}{w}} are fully expanded into their constituent character transitions.

\vspace{-0.05cm}
\subsubsection{Absolute Values \vs Variance When Interpreting Performance on Functional and Non-Functional Properties}
\label{sec:metrics-AssessingNonFunctional}

Using \bench and the metrics in~\cref{tab:evaluation-metrics}, we can compare the performance of regex composition methods.
For functional properties, it is clear that higher accuracy and lower generation time are preferable.
We assume that higher semantic similarity is desirable, \ie that the engineer's regex is a reliable ground truth. %

For the non-functional properties, no particular value is intrinsically better.
The preferred value depends on an engineer's preferences and their context.
For example, some engineers may prefer more verbose regexes or those that use simpler constructs, while others may prefer concise (even cryptic) synonyms.
Some may favor strict regexes, others more general ones.
Guided by the qualitative work of Michael \etal~\cite{michaelRegexesAreHard2019} (see also~\cref{subsec:requirements}),
we posit that engineers want \emph{variety}: a range of candidates allowing them to weigh different non-functional dimensions for their needs. %

We surveyed software engineers to substantiate this perspective for the non-functional properties.

\paragraph{Survey design}
Under IRB oversight, we designed a survey to probe engineers' perspectives on the value of variety in regex composition. %
Subjects consented and were screened on regex competence using questions like Chapman \etals~\cite{chapmanExploringRegexComprehension2017}. %
We then familiarized the subjects with our context by describing regex composition tools, showed a screenshot of \toolname, and primed them with our question:
 ``\textit{Often, there are dozens or hundreds of candidate regex solutions, and we are looking for ways to optimize the presentation of solutions to maximize usefulness. One idea we have is that developers may want to see solutions with a lot of variety.}''

After this, subjects completed 4 regex composition tasks selected from the most popular regexes from RegExLib.
In each task, subjects first selected their preferred regex from a list.
Then, they compared two sets of candidate solutions and chose which they would find more helpful (\cref{fig:survey-question}).
These sets were constructed so that one was high-variance and one was low-variance in the candidates' semantic similarity (``strictness'') to the known ground truth, randomly ordered.
Finally, subjects answered questions like ``\textit{Would seeing the sets lead you to reconsider your choice?}'' and whether greater variance helped.
We also collected demographic information.

\parpic[r]{%
  \begin{minipage}{0.45\textwidth}
    \centering
    \includegraphics[width=0.99\linewidth]{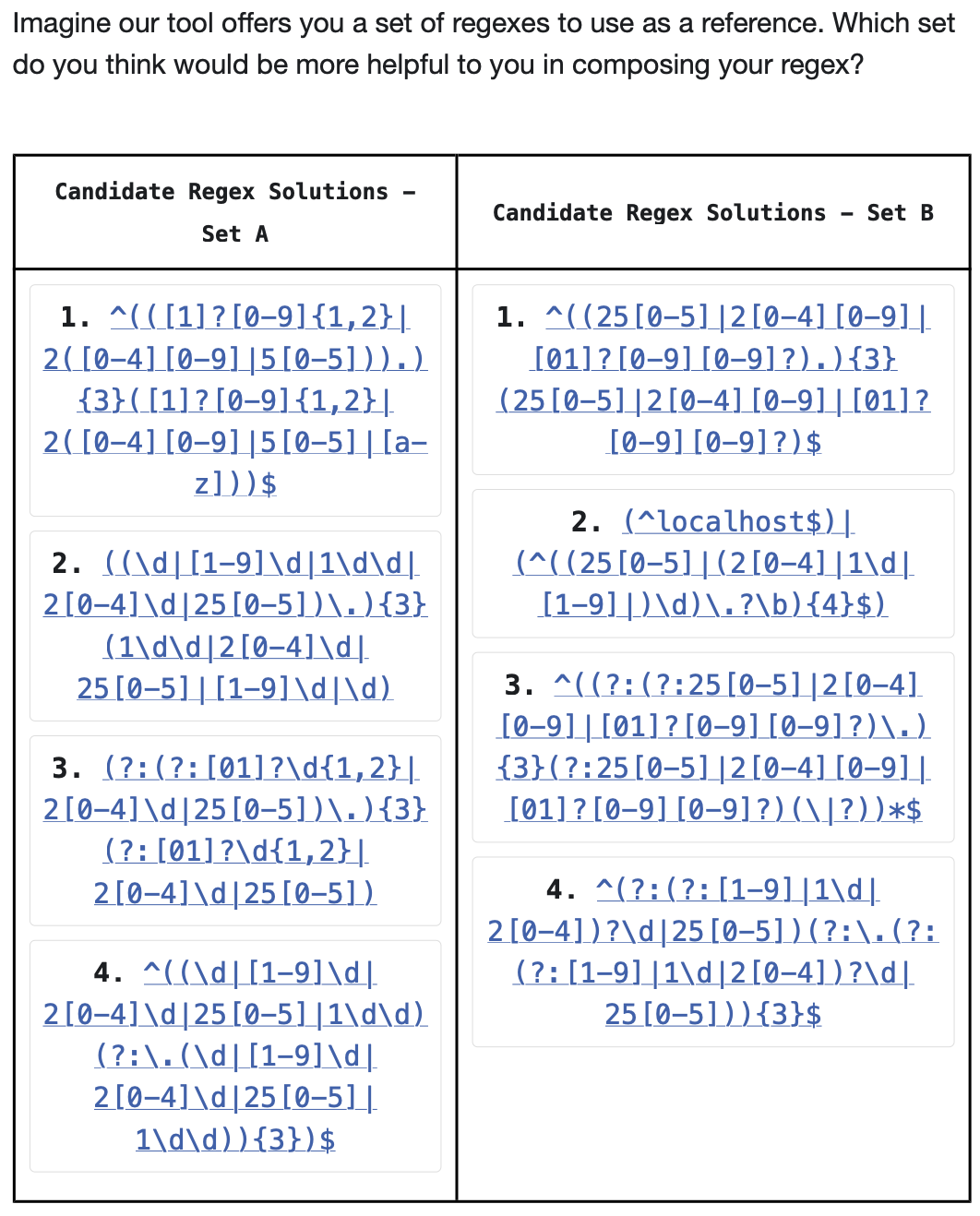}
    \captionof{figure}{Question prompts subjects to assess which \textit{candidate set} would be more useful.}
    \label{fig:survey-question}
  \end{minipage}%
}

We followed best practices such as those of Pfleeger \& Kitchenham~\cite{pfleeger2001principles}.
We refined the instrument over several weeks within the research team, piloted it with colleagues in our department, and trialed it on the Prolific platform.
The primary changes we made were:
  (1) improve face validity~\cite{thomas1992face} by being explicit about the context and our question about variety;
  and
  (2) ask about their preferred regex before asking which set they would rather consider, which helped them avoid choosing the set containing their preferred regex.
These changes yielded more detailed feedback from the final subjects on this topic.

\paragraph{Subject recruitment}
Subjects were recruited from the Prolific platform.
We selected for subjects with a job in software development or IT, with a non-EU location to avoid GDPR complications.
Subjects who passed the screening and completed the instrument received a \$8 incentive.
Prolific provides a reasonably spam-free platform, but we validated responses for internal consistency.

\paragraph{Results}
We received 38 responses, of which 32 passed the screening questions.
The median subject wrote a regex in the past month, self-reports intermediate regex skill, works for a company of 101-999 employees, and has 3--5 years of engineering experience.

Of these 32 subjects:
  23 of 32 agreed that seeing candidates with a range of strictness helped them choose;
  and
  about two-thirds of these preferred a wider range of strictness (\ie \textit{more variance}). %

We acknowledge the limited statistical power of this survey, but the results quantitatively support the earlier qualitative evidence of Michael \etal~\cite{michaelRegexesAreHard2019} about engineers' core challenge in regex composition.
In our analysis in~\cref{sec:results}, we therefore prefer a larger \textit{variance in absolute values} for the non-functional properties. Meanwhile, we use simple \textit{absolute values} for the functional properties.

We report the absolute values as well to accommodate alternative interpretations.

The survey instrument and complete set of responses are available in our \hyperref[sec:data-availability]{artifact}.

\subsection{Sampling \bench} \label{subsec:experimental-data}

For our experiments, we selected representative samples from \bench to balance computational, time, and monetary costs of evaluating regex composition strategies at large-scale, while preserving statistical validity.
We employed a stratified random sampling approach~\cite{parsonsStratifiedSampling2017} to account for the heterogeneous attributes of \bench.

Our sampling procedure involved several steps.
First, we pre-processed \bench to retain tasks with at least one positive and one negative example in their test suites.
Second, we excluded any task whose ground truth regex was incompatible with Python's \texttt{re} module.
Next, we annotated each remaining task with three attributes to serve as stratification variables:

\begin{itemize}
    \item \textit{Source}: Whether the test suite originated from open-source software projects \vs RegExLib,
    \item \textit{Test count}: The total number of positive and negative example strings in the test suite.
    \item \textit{Positive example ratio}: The proportion of positive examples within the test suite.
\end{itemize}

We then stratified along these three dimensions: source (2 levels), test count (partitioned into 4 strata by the quartiles of the overall test count distribution: 4 levels), and positive example ratio (partitioned into 4 strata by the quartiles of the overall ratio distribution: 4 levels), for 32 strata.

With these strata, we determined the overall sample size using Cochran's method~\cite{Cochran1977-wd}.
We used Neyman allocation~\cite{neymanTwoDifferentAspects1934} to proportionally distribute this total sample size across the strata, drawing a random set of regex composition tasks from each without replacement.

This procedure resulted in two disjoint sets:
    an \ul{evaluation set} of 374 tasks, sampled to provide 95\% confidence with a $\pm$5\% margin of error, which we used for our primary analyses;
    and
    an \ul{ablation set} of 68 tasks, sampled to provide 90\% confidence with a $\pm$10\% margin of error, which we used for tuning the parameters of the regex composition strategies.

\subsection{Regex Composition Method Selection and Configuration}
\label{sec:ExperimentSetup-ComparisonTargets}

Next, we select the representative techniques for each regex composition method and describe how we configured each technique.

\subsubsection{Large Language Models}
\paragraph{Model Selection} To evaluate the state-of-the-art LLM-based regex generation, we selected both proprietary and open-source models. %
\cref{tab:llm-models} shows these models.
\vspace{3mm}

\begin{enumerate}
\parpic[r]{%
\begin{minipage}{0.50\textwidth}
  \tiny
  \centering
  \captionof{table}{
  List of LLMs used in our evaluation, based on programming proficiency (top) and popularity (bottom).
  Proprietary LLMs have parameter counts estimated in the trillions~\cite{nasutoLearningTopicNot2025}.
  Proprietary models use vendors' cloud-based services (\Bicon[scale=0.7]{908}).
  Large models need servers (\Bicon[scale=0.7]{1406}).
  Smaller models fit on a developer workstation (\Bicon[scale=0.7]{1557}).
  }
  \label{tab:llm-models}
  \vspace{-1mm}
  \begin{tabular}{llccc}
    \toprule
    \textbf{Vendor} & \textbf{Model Name} & \textbf{Params.} & \textbf{Weights} & \textbf{Hw. Reqs.} \\
    \midrule
    OpenAI          & GPT-5               & \textbf{?}                   & Prop. & \Bicon[scale=0.7]{908} \\
    OpenAI          & GPT-5 mini          & \textbf{?}                   & Prop. & \Bicon[scale=0.7]{908} \\
    OpenAI          & o3                  & \textbf{?}                   & Prop. & \Bicon[scale=0.7]{908} \\
    xAI             & Grok 4              & \textbf{?}                   & Prop. & \Bicon[scale=0.7]{908} \\
    Alibaba Cloud   & qwen3-coder         & 480\ B                     & Open  & \Bicon[scale=0.7]{1406} \\
    DeepSeek AI     & deepseek-chat-v3.1  & 685\ B                     & Open  & \Bicon[scale=0.7]{1406} \\
    OpenAI          & gpt-oss-120b        & 120\ B                     & Open  & \Bicon[scale=0.7]{1406} \\
    OpenAI          & gpt-oss-20b         & 20\ B                      & Open  & \Bicon[scale=0.7]{1557} \\
    \midrule
    Anthropic       & Claude Sonnet 4     & \textbf{?}                   & Prop. & \Bicon[scale=0.7]{908} \\
    Google          & Gemini 2.5 Pro      & \textbf{?}                   & Prop. & \Bicon[scale=0.7]{908} \\
    Google          & Gemini 2.5 Flash    & \textbf{?}                   & Prop. & \Bicon[scale=0.7]{908} \\
    Google          & Gemini 2.0 Flash    & \textbf{?}                   & Prop. & \Bicon[scale=0.7]{908} \\
    DeepSeek AI     & deepseek-chat-v3    & 685\ B                   & Open  & \Bicon[scale=0.7]{1406} \\
    DeepSeek AI     & deepseek-r1         & 685\ B                     & Open  & \Bicon[scale=0.7]{1406} \\
    Alibaba Cloud   & qwen3-30b-a3        & 30\ B                      & Open  & \Bicon[scale=0.7]{1557} \\
    Mistral AI      & mistral-nemo        & 12\ B                      & Open  & \Bicon[scale=0.7]{1557} \\
    \bottomrule
  \end{tabular}
\end{minipage}}
    \item \textit{Models Selected for Their Programming Proficiency:} We used the best models on two major programming benchmarks as of August 2025: LiveCodeBench~\cite{jain2024livecodebench} and SWE-bench~\cite{jimenez2023swe}.
    From each benchmark, we selected four models: the best two proprietary and open-source models. %
    This yielded
      (proprietary)
      GPT-5 mini, o3, GPT-5, and Grok 4;
      and
      (open-source)
      gpt-oss-120b, gpt-oss-20b, qwen3-coder and deepseek-chat-v3.1.
    \item \textit{Models Selected for Their Popularity:}
      For another view, we also selected LLMs by usage statistics.
    Using public data from OpenRouter~\cite{openrouter_rankings} as of August 2025, we identified the eight most popular models, excluding those already selected for their programming proficiency. %
    These were:
      (proprietary)
      Claude Sonnet 4, Gemini 2.5 Flash, Gemini 2.0 Flash, and Gemini 2.5 Pro,
      and
      (open-source)
      deepseek-chat-v3, deepseek-r1, qwen3-30b-a3, and mistral-nemo.
\end{enumerate}
\vspace{4mm}
\paragraph{Configuration}
Prior work has established that clear problem statements with structured input-output examples enhance the accuracy of LLM-generated regexes~\cite{siddiqUnderstandingRegularExpression2024,siddiqReGExDoS2024}.
We began with the prompt from those previous works: \textit{``[...]. Generate a regex for this description. Matching strings: [...]. Non-matching strings: [...].''}.
However, we were skeptical of the relatively low initial success rate of that prompt (reported initial success on gpt-3.5 of 33\%, and only 44\% after 3 iterations).
Following OpenAI's guide~\cite{openai-prompt-engineering-guide}, we expanded this baseline prompt with more detailed instructions and contextual information.
Our full prompt incorporated strategies such as
  persona and a detailed response schema.

To assess our enhanced prompt, we used the gpt-oss-120b model to generate regexes for the tasks in our ablation set.
For each task, we generated a single candidate regex, with up to three retries that added failure feedback.
With the baseline prompt, the model solved the task on the first attempt 94.12\% of the time and within three attempts for 98.53\% of tasks, yielding an average accuracy of 92.62\% per generated regex.
In contrast, our enhanced prompt solved the task on the first attempt 95.59\% of the time and succeeded on all remaining tasks within three attempts (a 100.00\% success rate), with the average accuracy per regex increasing to 98.26\%.
To assess the risk of overfitting to one specific model, we also tested both prompts on gpt-oss-20b (workstation-grade) and found similar performance trends.

After identifying an improved prompt, we optimized two parameters:
  $K$, the number of alternative regex candidates generated per query;
  and $L$, the number of few-shot examples included in the prompt. %
We performed a grid search over the parameter space $K \in \{1,2,4,8\}$ and $L \in \{0,1,2,4,8\}$.
The best parameters were $K$=2 and $L$=0, which improved the per-regex accuracy by 1.14\% even though the task success rate remained constant.
We used this prompt and configuration in all experiments.
We set the temperature (or equivalent parameter) to 0 to increase determinism and replicability.
Our full prompt and the details of this parameter tuning study are in our \hyperref[sec:data-availability]{artifact}.

\subsubsection{Formal Regex Synthesis}
\paragraph{Synthesizer Selection}
We selected synthesizers based on three criteria:
  (i) the availability of a public artifact with clear execution instructions,\footnote{When an artifact was unavailable, we contacted some authors to request access, but they declined.}
  (ii) the ability to operate on positive/negative example strings of arbitrary alphabets (\ie no multimodal or NLP-based approaches, as our benchmark does not encompass task descriptions),
  and
  (iii) support for non-interactive execution to enable a large-scale analysis.
From the literature, three state-of-the-art synthesizers met our criteria:
  RFixer~\cite{panAutomaticRepairRegular2019},
  \textsc{Forest}~\cite{ferreiraFORESTInteractiveMultitree2021},
  and
  \textsc{RegexPlus}~\cite{pertseva2022regexPlus}.
The full rationale for our inclusion/exclusion criteria, including a list of synthesizers that were considered but ultimately excluded from our study, is detailed in our artifact.

\paragraph{Configuration}
To select the best performing synthesizer, we evaluated each selected synthesizer on the ablation dataset using its default parameter configuration.
Results are in~\cref{fig:ablation-synth}.
Based on this evaluation, we used RFixer in the subsequent experiments.

\subsubsection{Regex Reuse}

\paragraph{System Selection}
We implemented \toolname as described in~\cref{sec:reuse-by-example}.
No optimization was performed: it is a loop over Python regex engine queries for each candidate in the database.
For each task, we subtract the solution regex from \db prior to running the query.

\begin{figure}[ht]
  \centering
  \begin{subfigure}[b]{0.48\linewidth}
    \centering
    \includegraphics[width=1.54in,height=1.15in]{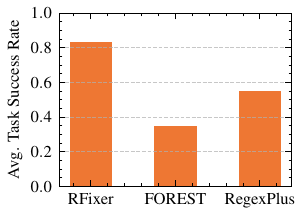}
    \caption{Synthesizer performance on ablation dataset. Based on its superior success rate, we selected RFixer for our main evaluation.}
    \label{fig:ablation-synth}
  \end{subfigure}
  \hfill
  \begin{subfigure}[b]{0.48\linewidth}
    \centering
    \includegraphics[width=1.54in,height=1.15in]{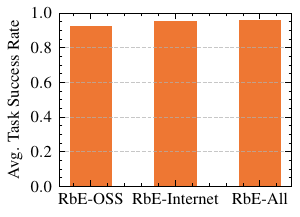}
    \caption{Regex reuse configurations on ablation dataset. We opted to use the RbE-All, which queries regexes from all sources, for our subsequent experiments.}
    \label{fig:ablation-RegexReuse}
  \end{subfigure}
  \caption{
  Ablation results for (a) formal synthesizers, and (b) reuse database configuration.
  }
  \label{fig:ablation-summary}
  \Description{}
\end{figure}

\paragraph{Configuration}
The primary configuration for an instance of \toolname is the underlying set of regexes that are queried.
The larger the database, the more potential candidates, but the more costly each query.
\db includes regexes from several kinds of sources (\cref{tab:db-stats}).
We ablated performance in three modes:
  regexes from open-source software projects only,
  regexes from internet sources only,
  and
  all.
\cref{fig:ablation-RegexReuse} shows the result on the ablation set: reuse performance is comparable regardless of the data source, but merging all sources does yield some benefit.
Based on this result, we used \toolname with the full \db in the subsequent experiments.

\section{Results}
\label{sec:results}

\subsection{Illustrative Comparison}

We begin by illustrating \bench and the results of different regex composition strategies on some standard tasks.
\cref{tab:regex-examples} presents regexes generated by the best performing
  LLM method (GPT-5),
  formal synthesis method (RFixer),
  and
  \toolname for two common pattern matching tasks:
  email validation\footnote{Email ground truth: \detokenize{^[-!#\%&'*+\\/0-9=?A-Z^_a-z`{|}~](\.?[-!#$\%&'*+\\/0-9=?A-Z^_a-z`{|}~])*@[a-zA-Z0-9](-*\.?[a-zA-Z0-9])*\.[a-zA-Z](-?[a-zA-Z0-9])+$.
  This task contains 7 positive and 6 negative strings}
}
  and
  IPv4 loopback address validation.\footnote{IPv4 ground truth: \detokenize{^127\.([0-9]{1,3})\.([0-9]{1,3})\.([0-9]{1,3})$}. This task contains 4 positive and 9 negative strings.}

These examples illustrate the challenge of the \bench tasks.
The ground truth regexes are complex, and the associated test suites do not fully specify the behavior.
The LLM approach produces RFC-compliant regexes.
The formal synthesis approach, RFixer, times out or overfits to the examples.
The solutions from reuse appear similar to the ground truth.
This is an expected property of the LLM and reuse-by-example approaches compared to RFixer: because I/O examples are a projection of the full behavior, one needs domain knowledge to produce a satisfying solution.
An LLM has this, and I/O examples can be used to retrieve regexes with comparable behavior.

\begin{table}[ht]
\centering
\scriptsize
\caption{Comparison of regexes produced by each composition method.}
\vspace{-0.1cm}
\label{tab:regex-examples}
\begin{tabularx}{\textwidth}{@{} l >{\raggedright\arraybackslash}X >{\raggedright\arraybackslash}X >{\raggedright\arraybackslash}X @{}}
\toprule
\textbf{Task} & \textbf{GPT-5} & \textbf{RFixer} & \textbf{\toolname} \\
\midrule
Email validation & \textasciicircum(?=.\{1,254\}\$)(?=.\{1,64\}@)[A-Za-z0-9]+(?:[\.\_\%\+\-][A-Za-z0-9]+)*@(?:[A-Za-z0-9](?:[A-Za-z0-9-]\{0,61\}[A-Za-z0-9])?\.)+[A-Za-z]\{2,63\}\$ & N/A (Timeout) & \textasciicircum\textbackslash w+([.-]?\textbackslash w+)*@\textbackslash w+([.-]?\textbackslash w+)*(\.\textbackslash w\{2,\})+\$ \\
\addlinespace %
IPv4 loopback addresses & \textasciicircum127\.(?:25[0-5]|2[0-4]\textbackslash d|1\textbackslash d\{2\}|\textbackslash d\{1,2\})\.(?:25[0-5]|2[0-4]\textbackslash d|1\textbackslash d\{2\}|\textbackslash d\{1,2\})\.(?:25[0-5]|2[0-4]\textbackslash d|1\textbackslash d\{2\}|\textbackslash d\{1,2\})\$ & ((12.[570]])\{3\})\{3,5\} & \textasciicircum127(?:\.(?:25[0-5]|2[0-4][\textbackslash d]|[01]?[\textbackslash d][\textbackslash d]?))\{3\}\$ \\
\bottomrule
\end{tabularx}
\end{table}

Next, we report the performance of each strategy on each metric in turn.
For the LLM-based approach, we present results only for the best-performing model from each deployment category (commercial/cloud, open/server, and open/workstation), which we selected based on preliminary accuracy evaluations.
Due to space limitations, analyses of other functional and non-functional properties are reported for these three representative models.
The accuracy results for all other LLMs are available in our \hyperref[sec:data-availability]{artifact}.

\subsection{Functional Properties}
\label{subsec:results-functional}

\begin{figure}[t]
  \centering

  \begin{subfigure}[b]{0.49\linewidth}
    \centering
    \includegraphics[width=2.68in,height=1.725in]{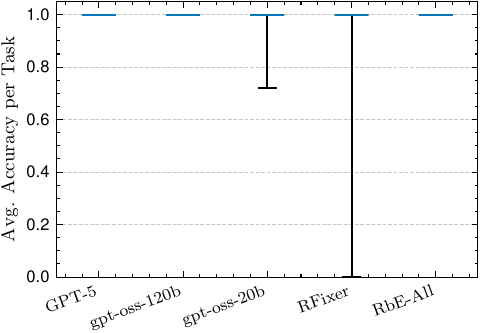}
    \caption{Accuracy}
    \Description{}
    \label{fig:AccuracyPerTask}
  \end{subfigure}
  \hfill
  \begin{subfigure}[b]{0.49\linewidth}
    \centering
    \includegraphics[width=2.68in,height=1.725in]{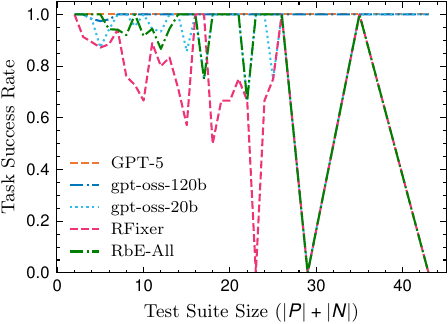}
    \caption{Accuracy by test suite size}
    \Description{}
    \label{fig:AccuracyBySuiteSize}
  \end{subfigure}

  \vspace{1ex}

  \begin{subfigure}[b]{0.49\linewidth}
    \centering
    \includegraphics[width=2.68in,height=1.725in]{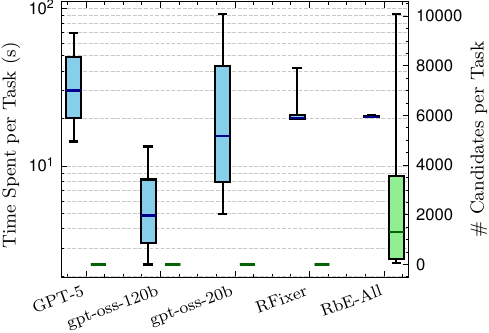}
    \caption{Generation time and \# candidates}
    \Description{}
    \label{fig:generationTime}
  \end{subfigure}
  \hfill
  \begin{subfigure}[b]{0.49\linewidth}
    \centering
    \includegraphics[width=2.68in,height=1.725in]{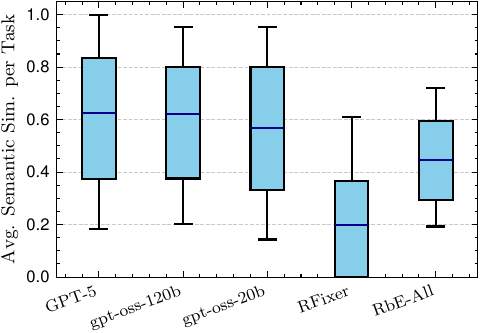}
    \caption{Semantic similarity}
    \Description{}
    \label{fig:SemanticSimilarity}
  \end{subfigure}

  \caption{Comparison of regex composition strategies across functional metrics.
  For the box plots, the distributions represent: (a) the average accuracy across all candidates generated for a given task, (c) the total generation time and number of candidates retrieved per task, and (d) the average semantic similarity of all candidates generated for a given task to the ground-truth solution.
  In subplot (c), the blue box plots correspond to the generation time (left log-scale y-axis), while the green box plots represent the number of candidates found (right y-axis).
  The line chart in (b) plots the success rate (\ie being able to generate at least one candidate that satisfies the given task) of each method as a function of the test suite size.
  In all box plots, whiskers denote the 10th and 90th percentiles of the distribution.}
  \label{fig:FunctionalProperties}
  \vspace{-4mm}
\end{figure}

We begin with \textit{accuracy}, the most basic measure of each composition strategy.
As seen in~\cref{fig:AccuracyPerTask}, the two larger LLMs and \toolname yield candidates that achieve perfect accuracy on more than 90\% of tasks.
They almost always produce at least one perfect solution.
The workstation-grade LLM (gpt-oss-20b) struggles more, and RFixer fails $\sim$25\% of the time (consistent with~\cref{fig:ablation-synth}).

We were curious about any \textit{relationship between task difficulty and the performance} of the regex composition strategies.
\cref{fig:AccuracyBySuiteSize} plots each method's success rate, defined as generating at least one candidate that satisfies the task, across test suite sizes.
The LLMs are consistently successful, while more complex tasks are more challenging for RFixer and \toolname.

We also assessed \textit{how many distinct solutions} one can obtain with \toolname.
The green bars in~\cref{fig:generationTime} indicate that the median task in \bench has $\sim$500 distinct candidate solutions in \db.
For the other methods, our prompt fixed the number of LLM candidates to two, and RFixer is designed to produce one result.

Turning our attention to \textit{time spent}, examine~\cref{fig:generationTime}.
Note the log scale on the left axis.
For the commercial LLM, the need for multiple iterations over a web API makes the median generation time roughly 30 seconds.
Using a local GPU (though a top-of-the-line H100), LLM generation was much faster with the server-grade LLM (median $\sim$3 seconds), while the workstation-grade LLM was slower relative to the server-grade setup (median $\sim$15 seconds) because of the increased need for iterative feedback.
Meanwhile, \toolname's stable algorithm takes $\sim$20 seconds per task,
and RFixer occasionally takes longer.

Finally, in~\cref{fig:SemanticSimilarity}, we consider the semantic similarity between the ground truth and the candidates proposed by each strategy. LLMs and \toolname produce candidates of comparable similarity to the ground truth.
RFixer's are less similar, perhaps because it is designed to produce a conservative solution.

\subsection{Non-Functional Properties}
\label{subsec:results-nonfunctional}

\begin{figure}[t]
  \centering
  \begin{subfigure}[b]{0.49\linewidth}
    \centering
    \includegraphics[width=2.68in,height=1.725in]{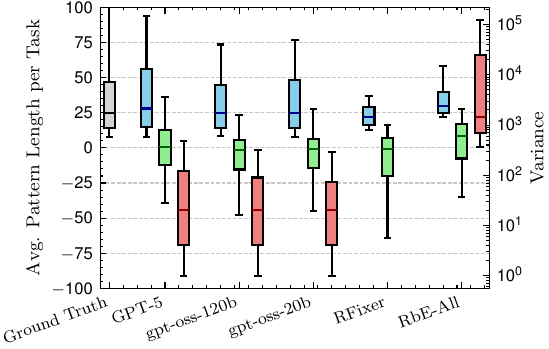}
    \caption{Pattern length}
    \Description{}
    \label{fig:pattern-length}
  \end{subfigure}
  \hfill
  \begin{subfigure}[b]{0.49\linewidth}
    \centering
    \includegraphics[width=2.68in,height=1.725in]{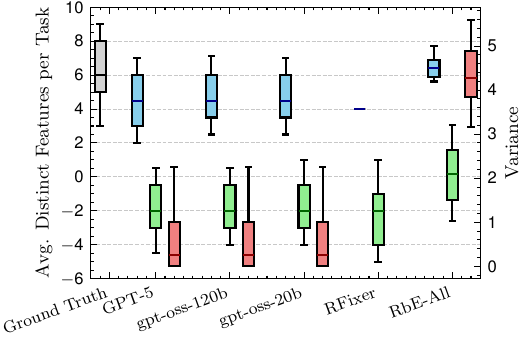}
    \caption{Distinct regex features used}
    \Description{}
    \label{fig:distinct-features}
  \end{subfigure}

  \vspace{1ex}

  \begin{subfigure}[b]{0.49\linewidth}
    \centering
    \includegraphics[width=2.68in,height=1.725in]{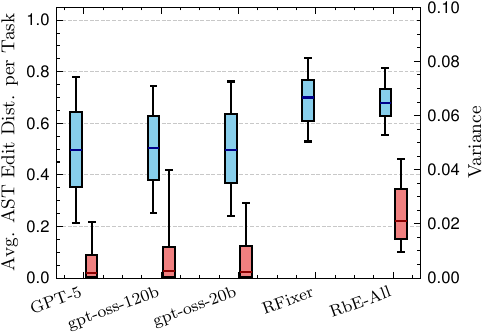}
    \caption{AST edit distance}
    \Description{}
    \label{fig:syntactic-sim}
  \end{subfigure}
  \hfill
  \begin{subfigure}[b]{0.49\linewidth}
    \centering
    \includegraphics[width=2.68in,height=1.725in]{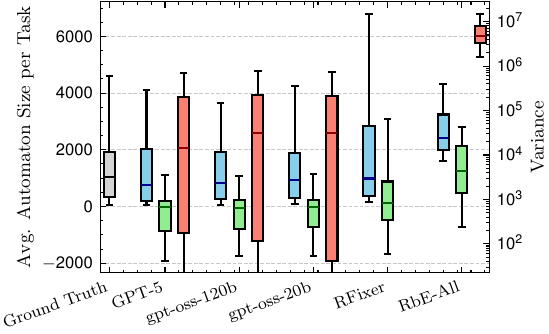}
    \caption{Automaton size}
    \Description{}
    \label{fig:automaton-size}
  \end{subfigure}

  \caption{Comparison of regex composition strategies across the non-functional metrics.
  In each subplot, the gray plots show the distribution for the ground-truth solutions.
  For each composition method,
  the blue plots show the distribution of the metric's absolute value, averaged across all candidates generated per task;
  the green plots show the distribution of the difference of these averages relative to the corresponding ground truth solutions (effectively, the blue plot's value minus the gray plot's value for each task);
  and the red plots show the per-task variance across the set of yielded candidates (corresponding to the right y-axis, log scale for (a) and (d)).
  For all box plots, the whiskers denote the 10th and 90th percentiles of the distribution.}
  \label{fig:NonFunctionalProperties}
  \vspace{-4mm}
\end{figure}

On \textit{pattern length}, refer to~\cref{fig:pattern-length}.
All approaches produced candidates of comparable length to the ground truth.
\toolname offered far more variety in the length of solutions (red bars).

For the number of \textit{distinct regex features used}, examine~\cref{fig:distinct-features}.
Here, the human-authored regexes (ground truth and \toolname) both used more features than the LLM-based approaches.
Note that RFixer has a constrained feature set; thus, it uses a constant number of features.
Again, \toolname offers much greater variance in candidates than the others.

Regarding \textit{syntactic similarity}, see~\cref{fig:syntactic-sim}.
In this case, the LLMs produce solutions that are more syntactically similar to the ground truth (smaller edit distances), while RFixer and \toolname are both somewhat further.
Once more, \toolname offers greater variance than the others, though in this case the absolute value of the variance is small.

Finally, for \textit{automaton size}, examine~\cref{fig:automaton-size}.
Here, there is greater overlap in the median automaton sizes across all approaches, though \toolname again varies more.

\section{Discussion and Future Work}
\label{sec:Discussion}

\subsection{Is Reuse All You Need?}

Summarizing our results on the functional metrics (\cref{fig:FunctionalProperties}):
  all approaches
    solve at least half of the regex composition tasks, with over 90\% success by the LLM and \toolname methods (\cref{fig:AccuracyPerTask}),
  and
  they typically take less than a minute to run.
However, in terms of the non-functional metrics, there is no comparison---the \toolname offers the widest variety of candidates, which is the property that our analysis and user study suggest matters (\cref{sec:metrics-AssessingNonFunctional}).

We believe these data show that for the \bench tasks (with biases as discussed in~\cref{sec:validity-threats}), reusing regexes offers a favorable cost-benefit tradeoff.
\toolname offers a diversity of accurate solutions at low relative cost.
In light of the astounding median of 500 candidates per task (\cref{fig:generationTime}), our interpretation is that most regex tasks have already been solved by someone else.
The challenge lies not in solving new regex problems but in choosing among other engineers' solutions.
This outcome is different from several other recent works that addressed more complex tasks, \eg automated patch generation~\cite{xia2023automated} and solving security tasks~\cite{hamer2024just}, in which LLMs substantially outperformed existing techniques.

We acknowledge an alternative perspective:
  perhaps the wide variety of regex candidates derived from \toolname occurs because engineers have made hundreds of different mistakes,
  and LLMs produce relatively unvarying solutions because they are referencing formal specifications such as the RFCs describing email and IP addresses.
Further empirical studies might assess the degree of (non)-adherence to such specifications, which would shed more light on the need for many variations.

\subsection{Future Work}

\subsubsection{Open Problems in Regex Composition}

If reuse is all you need for regex composition, a natural next step is to refine and evaluate this strategy with users.
The challenge may lie in determining which solutions (among hundreds) to present to a user, and how to help them choose.
Prior work has shown the difficulty of comprehending and comparing regexes~\cite{chapmanExploringRegexComprehension2017}, but no validated formulae exist to guide a sorting approach.
Also, our \toolname prototype is not optimized in terms of time efficiency (\cref{fig:generationTime}), and might be improved through clustering or automata merging.

When engineers apply regexes to novel problems or in niche domains, we expect reuse would be less useful.
On such hard tasks, existing regexes might be split or combined to solve new problems, a tactic we did not explore in our present prototype.
In this setting, we conjecture that a hybrid approach---combining reuse and LLMs, and possibly formal synthesis---might be suitable.
A benchmark will need to be established in order to assess such an approach.
We note that using LLMs for such techniques introduces concerns about provenance and potential licensing constraints~\cite{wang2025insights}, which can be more easily mitigated by reuse and formal synthesis methods.

\subsubsection{Applications of \db and \bench}

The \db corpus provides an updated dataset of engineer-made regexes that can be used to fine-tune large language models for improved regex generation and understanding.
Beyond model training, \db also supports the development of new regex engineering tools, such as input generators and visualization platforms, by reflecting authentic usage patterns and edge cases encountered in practice.

The \bench benchmark enables rigorous evaluation and comparison of regex composition algorithms, supplying realistic, example-driven tasks that mirror actual developer workflows.
Although the test suites extracted from real-world projects are often incomplete~\cite{wangHowWellAre2018}, they offer valuable insight into the kinds of inputs engineers use in practice.
This opens the possibility of learning effective test suite construction strategies from empirical data, rather than relying solely on exhaustive input generation.
\bench may also be a resource for the regex formal synthesis community, whose current datasets (as noted in~\cref{subsec:regex-composition-benchmarks}) are generally unrealistic.

\section{Threats to Validity}
\label{sec:validity-threats}

We discuss construct, internal, and external threats to validity.
Following Verdecchia \etals guidance~\cite{verdecchia2023threats}, here we
focus on substantive threats that may affect our results.

\textbf{Construct validity}
assesses whether the chosen measures and instruments correctly operationalize the theoretical concept they intend to capture.
We introduced two constructs in this work.
First, we developed the \toolname system for regex composition, which we grounded in the programming by example framework.
Second, we introduced the use of variance as a way to interpret measures of non-functional properties for the range of solutions offered by each regex composition strategy.
Our approaches to each were based on prior qualitative studies of software engineers' approaches to regex composition and assessment, and for the latter we validated it with a small user study (\cref{sec:metrics-nonFunctional}).
The other measures in this study are reused or adapted from prior work.

\textbf{Internal validity}
refers to the extent to which a study establishes a causal relationship between the independent and dependent variables.
Our work evaluates no cause-effect relationships.
However, based on our results, we proposed the following conjecture:
  For many regex use cases, reuse is all you need.
Our results indicate that this conjecture is worth investigation.

\textbf{External validity}
  refers to the extent to which the results of our study can be generalized.
Like any empirical work, our results are constrained by the realities of large-scale measurement.
We highlight the most critical threats.
First, our study considers state-of-art approaches for formal synthesis
and LLMs, restricting the input mode to only positive and negative examples.
Different results might occur if we considered a broader range of inputs (\eg a draft or defective regex, or a natural language description).
However, as both \toolname and LLM approaches achieved near-perfect composition results, we see little room for improvement.
Second, our construction of \bench and our evaluation rely on the Python regex semantics.
There are variations by programming language, but Davis \etal reported that for real-world regexes these are rare~\cite{davisWhyArenRegular2019b}.
Third, our results may not generalize to regex composition tasks not present in our dataset.
Davis \etal showed that regexes mined from open-source software projects tend to fall into categories associated with web servers (\eg emails, file names, HTML, URL, user-agent strings)~\cite{davisImpactRegularExpression2018a}.
Regexes and I/O examples for tasks such as intrusion detection and antivirus scanning may not be represented enough in \db; prior measurements suggest that such regexes may be structurally different.
Beyond regexes, our results may not generalize to other software composition tasks (\eg SQL query generation; function reuse \vs synthesis).
Part of the reason regex composition has been such a longstanding target for program synthesis is that the problem is well constrained; this same property suggests that the problem may be amenable to \toolname in a way that more complex software structures are not.

\section{Conclusion}
We provided the first systematic comparison of regex composition strategies.
We leveraged a novel dataset of regex composition tasks mined from open-source software projects and internet sources.
We operationalized engineers' common regex reuse practices with \toolname, and compared this programming by example approach against state-of-the-art formal synthesis and LLM-based approaches.
We report that \toolname and LLM-based approaches both solve these tasks with near-perfect accuracy, while formal synthesis approaches struggle more often.
Meanwhile, \toolname is distinct in the variety of solutions it offers to engineers, along every dimension we considered (\eg length, features used, automaton size).
Given the substantial difference in operating costs, a refined reuse scheme may be preferable to LLMs for regex composition.

\section*{Data Availability}\label{sec:data-availability}

An artifact associated with this paper is available at \url{https://anonymous.4open.science/r/regex-evaluation-27D2}.

\ifANONYMOUS
\else
\begin{acks}
This work was supported in part by the \grantsponsor{YRXVL4JYCEF5}{National Science Foundation (NSF)}{} under grant
\#\grantnum{YRXVL4JYCEF5}{2135156}.
\end{acks}
\fi

\printbibliography

\end{document}